\documentclass[manuscript]{aastex}

\shorttitle{Coronal heating above the disks}
\shortauthors{Takasao et al.}

\usepackage{bm}
\usepackage{amsmath}
\usepackage{color}

\begin{document}


\title{A THEORETICAL MODEL OF X-RAY JETS FROM YOUNG STELLAR OBJECTS}


\author{Shinsuke Takasao\altaffilmark{1}}
\email{takasao@kwasan.kyoto-u.ac.jp}
\author{Takeru K. Suzuki\altaffilmark{1,2}}

\and

\author{Kazunari Shibata\altaffilmark{3}}


\altaffiltext{1}{Department of Physics, Nagoya University, Nagoya, Aichi 464-8602, Japan}
\altaffiltext{2}{School of Arts \& Sciences, University of Tokyo, 3-8-1, Komaba, Meguro, Tokyo, 153-8902}
\altaffiltext{3}{Kwasan and Hida Observatories, Kyoto University, Yamashina, Kyoto 607-8471, Japan}


\begin{abstract}
There is a subclass of the X-ray jets from young stellar objects which are heated very close to the footpoint of the jets, particularly DG Tau jets. Previous models attribute the strong heating to shocks in the jets. However, the mechanism that localizes the heating at the footpoint remains puzzling. We presented a different model of such X-ray jets, in which the disk atmosphere is magnetically heated. Our disk corona model is based on the so-called nanoflare model for the solar corona. We show that the magnetic heating near the disks can result in the formation of a hot corona with a temperature of $\gtrsim 10^6$~K even if the average field strength in the disk is moderately weak, $\gtrsim 1$ G. We determine the density and the temperature at the jet base by considering the energy balance between the heating and cooling. We derive the scaling relations of the mass loss rate and terminal velocity of jets. Our model is applied to the DG Tau jets. The observed temperature and estimated mass loss rate are consistent with the prediction of our model in the case of the disk magnetic field strength of $\sim20$~G and the heating region of $<0.1$~au. The derived scaling relation of the temperature of X-ray jets could be a useful tool to estimate the magnetic field strength. We also found that the jet X-ray can have a significant impact on the ionization degree near the disk surface and the dead-zone size.
\end{abstract}



\section{INTRODUCTION}
Bipolar outflows and jets are one of the most outstanding manifestations of the young stellar objects (YSOs). Outflows and jets are believed to play important roles in the stellar and disk evolution through the removal of mass and angular momentum from the system \citep{blandford1982,ustyugova1999,dullemond2007,frank2014}, which have motivated many theoretical studies. Our common understanding is that a magnetic field is essential for accelerating and collimating the flows.

Fast jets often produce X-rays, and the X-ray emission has been often attributed to the shock heating. The most prominent example is the Herbig-haro objects that contain shock-heated plasma at or near the bow shock regions \citep[e.g. HH2;][]{pravdo2001}. Since the shock temperature can be expressed as $T\approx 1.5\times 10^5 (v_{\rm shock}/100~{\rm km~s^{-1}})^2$~K, where $v_{\rm shock}$ is the shock jump speed \citep{raga2002}, the typical jet speed of a few hundreds~${\rm km~s^{-1}}$ \citep[e.g.][]{ray2007} can result in the formation of X-ray emitting MK plasma through the collision with the surrounding material. A distinct example is L1551 IRS~5, which shows X-ray emission near the jet base ($\sim$100~au from the central star, \citet{bally2003}). To account for the X-ray emission, different shock formation processes near the base and in the jet have been discussed \citep{bally2003,bonito2010}.

\citet{gudel2005,gudel2008} discovered a spatially resolved bipolar X-ray jet from DG Tau, one of the best-studied classical T-Tauri stars (CTTSs). {\it Chandra} high-resolution X-ray images revealed the jet axis is coincident with the optical jet axis. The apparent length is $\sim5^{\prime\prime}\approx 700$~au (the distance of 140~pc is assumed). The X-ray jets are different from a typical X-ray source HH X-ray jets which are often seen at or near the bow shock regions (e.g. HH2).
It is found that the jet can be followed down into the point spread function (PSF) of DG Tau ($<0.1-0.2^{\prime\prime}\approx 30$~au), suggesting the heating very close to the central star. The mass loss rate is estimated to be $1.3\times 10^{-11}M_\odot$~yr$^{-1}$ \citep{schneider2008}. The inner jet spectra are softer than the stellar spectra, with a best-fit electron temperature of $\sim$3~MK, while the hard component from the star has a temperature of $>$20~MK. The hard component is highly time variable, showing the flaring activity in the stellar corona. However, the soft X-ray emission from the inner jet or the jet base in the PSF remains steady and stationary over six years \citep{gudel2011}. This suggests that the X-ray jet may not be originated from the star and a continuous heating is operating near the jet base, although the pulsed jet scenario of \citet{bonito2010} may explain the apparent steady emission. The soft X-ray luminosity of the DG Tau X-ray jets is estimated to be $\sim 10^{28-29}$~erg~s$^{-1}$, which is not negligible compared to the stellar X-ray luminosity, $\sim 10^{30}$~erg~s$^{-1}$ \citep{gudel2008}. \citet{schneider2011} analyzed a similar soft X-ray source in the L1551 IRS 5 jet (see also \citet{favata2002,bally2003,bonito2011}). RY Tau and RW aur also show soft X-rays that could be related to jets \citep{skinner2014,skinner2016}.

Observations raise a fundamental question about the origin of the DG Tau X-ray jets. Although shock heating may be relevant \citep[e.g.][]{bonito2010}, a very large shock jump speed of 450~km~s$^{-1}$ is required to produce the observed 3~MK temperature. The jet of L1551 IRS5 showed a 500~km~s$^{-1}$ high-speed component near the outer X-ray emitting region, supporting the shock heating scenarios \citep{fridlund2005}, but not around the inner jet \citep{pyo2009}. Such a high velocity component is not always observed and has not been detected yet for the DG Tau jets \citep{lavalley2000,rodriguez2012}. Optical forbidden emission line and far-ultraviolet observations favor the presence of slower shocks with a speed of $50-100$~km~s$^{-1}$ for the DG Tau jets \citep{lavalley2000,schneider2013b}. Multiple shock heating may resolve the problem \citep{bonito2010,staff2010}, but the shock heating models based on (M)HD simulations need more development, because they do not self-consistently include key physics that identifies the injection speed, the time-variability, and the acceleration mechanisms of jets. The shock formation by the recollimation of stellar winds has been proposed \citep{gunther2014}, but it is unclear if the highly time-dependent stellar wind disturbed by the stellar flaring activity \citep[e.g.][]{schwenn2006} can form steady and stationary shocks. The gradual increase of the luminosity of the jet base reported by \citet{gudel2011} is inconsistent with this idea of the cooling remnants \citep{schneider2013b}. Further investigations on the heating process are necessary.

Large scale X-ray sources from jets can affect the protoplanetary disks. Soft X-rays can efficiently drive a photoevaporation wind from the disk \citep[e.g.][]{gorti2009,ercolano2009}. Vertically injected X-rays from jets can also raise the ionization degree of a large fraction of the protoplanetary disk, which can alter the accretion structure in the disk e.g. by inducing the magneto-rotational instability (MRI) or by suppressing the non-ideal magnetohydrodynamic (MHD) effects. The faint and fine X-ray structures like the DG Tau X-ray jets should be hard to detect because of attenuation, but there is a possibility that many YSOs have such X-ray sources. If this is the case, one needs to consider the effects when modeling the system.

The observable property of jets, such as the mass loss rate, depends on the acceleration mechanism. Therefore, one has to consider not only the heating but the acceleration mechanisms for better comparison between theoretical models and observations. It has been well-known that a magnetic field plays important roles in accelerating jets in different ways \citep{blandford1982,shibata1986,kudoh1995}. The gas pressure may become important in some cases \citep{takahara1989}. In this paper, we will consider the possible role of a magnetic field in heating the plasma at the base of jets, and determine the jet property on the basis of theories of jets.

\section{SPECIFYING DENSITY AND TEMPERATURE OF BASE OF MHD JETS}

The mass loss rate of jets, an important quantity that characterizes jets, is determined by the physical property at the jet base. A physically meaningful definition of the jet base is the slow point, where the slow point is the location where the jet velocity reaches the MHD slow-mode speed \citep{sakurai1985,pudritz1986,kudoh1997a}.
The energy balance and the mass loading across the jet base determine the physical property there. However, these processes have not been investigated in detail in the previous global jet models. We aim to specify the jet base condition.

\subsection{Basic Assumptions of Jet Base Heating}
The disk surface layer is expected to be unstable to MRI because of the sufficient ionization by cosmic rays and stellar irradiation \citep{turner2008}. The MRI-driven stretching and shearing motions will amplify magnetic energy in the disk. The amplified magnetic fields will eventually become buoyant, which will lead to the formation of rising magnetic loops (the Parker instability or magnetic buoyancy instability \citep{parker1966}). The magnetic energy supply by rising loops will result in the formation of a magnetically-dominated corona above the disk. The release of this magnetic energy affects the atmospheric structure.

\begin{figure}
\epsscale{.60}
\plotone{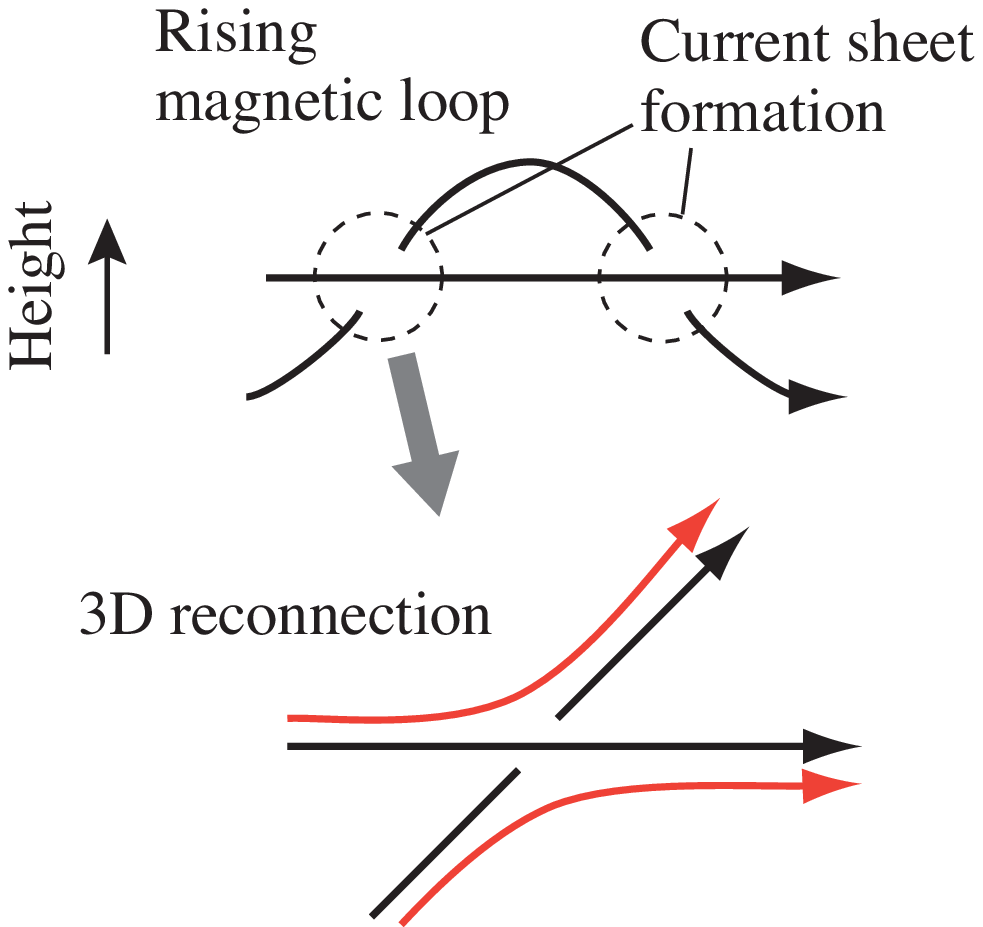}
\caption{Magnetic reconnection between a rising magnetic loop and an ambient azimuthal field. Current sheets are formed between them because of the steep change of the direction of magnetic fields. The current sheets are promising site for magnetic reconnection. As a result of magnetic reconnection, the magnetic field change its topology (red field lines are reconnected lines).\label{fig:rising_loop}}
\end{figure}

\citet{hirose2011} performed a 3D radiation-MHD simulation and suggested that magnetic energy dissipation in the upper disk occurs around the current sheets formed between rising loops and azimuthal ambient field lines. At such current sheets, magnetic reconnection will be a central energy release mechanism (see Figure~\ref{fig:rising_loop}). \citet{io2014} showed that magnetic heating can form a hot corona above a disk with the specific heat ratio of $>1.1$. Such a condition will be satisfied in the inner disk because the opacity is sufficiently high owing to the large density. For these reasons, we consider that magnetic heating will be important for determining the thermal structure. Here we focus on the heating by magnetic reconnection, although other magnetic heating processes such as the Alfv\'en wave heating may also contribute to the heating. This means that we utilize the idea of the nanoflare heating by \citet{parker1988}.

We first estimate the heating rate in the disk atmosphere. The successive formation of magnetic loops is a consequence that the escape and amplification of a magnetic field repeatedly occur from and in the disk, respectively. We take the escape timescale to be the timescale of the magnetic buoyancy, namely the growth timescale of the Parker instability, although there is a possibility that the amplification determines the escape timescale. The amplification timescale is similar to the rotational timescale because MRI or the rotational shear should be the major amplification mechanism. Since the former timescale is quite longer than the latter timescale, we consider that the energy injection timescale is governed by the growth of the Parker instability:
\begin{align}
\omega_{\rm P}\approx 0.3 v_A/H_{\rm disk} \sim 0.2\beta_{\rm disk}^{-1/2}\Omega,\label{eq:growth_rate}
\end{align}
where $v_{A}$ is the Alfv\'en speed in the disk, $H_{\rm disk}=\sqrt{2}a_{\rm disk}/\Omega$ is the pressure scale height, $\Omega$ is the Keplerian angular velocity, and $\beta_{\rm disk}$ is the plasma beta in the disk (the ratio of the gas pressure to the magnetic pressure). $a_{\rm disk}=\sqrt{\gamma R_g T_{\rm disk}}$ is the sound speed in the disk, where $R_g$ is the gas constant and $T_{\rm disk}$ is the disk temperature. 

Magnetic reconnection will release only a fraction of the magnetic energy supplied by the rising loops because the reconnection field lines will not be perfectly antiparallel (see Figure~\ref{fig:rising_loop}. Such magnetic reconnection is observed in the solar atmosphere: \citet{cirtain2013}). Therefore, using a constant nondimensional parameter $f$ ($0<f<1$), we express the heating rate per unit volume as
\begin{align}
E_{\rm H} =  f \frac{B_{\rm \phi}^2}{8\pi}\omega_{\rm P} \approx \frac{0.2f}{\gamma} \beta_{\rm disk}^{-3/2}\Omega a_{\rm disk}^2 \rho_{\rm disk}.
\end{align}
Considering MHD simulation results \citep[e.g.][]{hawley1995,matsumoto1995,sano2001}, we assume that the azimuthal component of the magnetic fields $B_\phi$ is commonly much greater than the poloidal component in the disk. 
The typical reconnection angle between the horizontal and vertical components should be neither very close to 0 nor very close to $\pi/2$.
For this reason we take $f=0.1$ as a fiducial value. The choice of $f=0.1$ means that the typical reconnection angle is approximately $\theta=\pi/5$ and the magnetic energy of the reconnection field is $B^2\sin^2{(\theta/2)}/8\pi \approx 0.1 B^2/8\pi$.

For a fully ionized and tenuous plasma such as the solar corona, the heat conduction will be the most dominant cooling process. We a priori assume that a hot, fully ionized corona is formed above the disk as a result of the magnetic heating. This assumption should be checked when we apply this model to a specific object. The conduction cooling rate per unit volume can be expressed as $\partial F_{\rm hc}/\partial s \approx \kappa_{\rm SP} T / L^2 = \kappa_0 T^{7/2}/L^2$, where $F_{\rm hc}$ is the heat conduction flux, $T$ is the temperature, and $s$ is the spatial coordinate along a field line. $\kappa_{\rm SP}=\kappa_0 T^{5/2}$ is the Spitzer conductivity \citep{spitzer1962}, where $\kappa_0$ is a constant and $\sim 10^{-6}$ in cgs units. $L$ is a typical length scale. We take the loop length as the length $L$, as in \citet{parker1988}, and the loop length as the wavelength of the Parker instability with the maximum growth rate, $\lambda_{\rm P}$. Then, we get
\begin{align}
\frac{\partial F_{\rm hc}}{\partial s} \approx \frac{\kappa_{\rm SP} T}{\lambda_{\rm P}^2} \approx \frac{\kappa_0 T^{7/2}}{\lambda_{\rm P}^2}.
\end{align}

\subsection{Scaling Relations of Physical Quantities of Jet Base}
We derive scaling relations of physical quantities of the jet base. Considering that hotter jets generally emanate from the region closer to the central star than cooler jets \citep[e.g.][]{bacciotti2000,schneider2013}, we expect that X-ray jets are originated from the inner disk region. As a typical inner disk model, we adopt the following viscous disk model. The disk temperature $T_{\rm disk}$ and density $\rho_{\rm disk}$ are written as
\begin{align}
T_{\rm disk} & = T_0 \left( \frac{r}{r_0}\right)^{-3/4} \label{eq:disk_temp}\\
\rho_{\rm disk} & = \rho_0 \left( \frac{r}{r_0}\right)^{-15/8}\label{eq:disk_rho}
\end{align}
where $r$ is the radius from the central star \citep{kusaka1970}. The Keplerian rotation ($\Omega \propto r^{-3/2}$) is assumed. The typical inner disk radius $r_0$, temperature $T_0$, density $\rho_0$, and stellar mass are taken from observations of DG Tau. \citet{gudel2007} estimated the stellar mass as $0.9-1.8M_\odot$, and therefore we take $1M_\odot$ as the fiducial stellar mass. On the basis of near-infrared observations \citep{akeson2005}, we take 0.1~au and $10^3$~K as an inner disk radius and the temperature, respectively. This radius is consistent with the corotation radius \citep{bouvier1993}. The typical density at the radius is estimated from the observed mass accretion rate. The density and the accretion rate are related by the relation $\dot{M}\approx 4\pi r_0 v_{\rm r} h \rho_0$, where $r_0=0.1$~au, $v_{\rm r}$ is the accretion velocity, $h$ is the pressure scale height at the radius $r_0$. In an $\alpha$-viscous disk model, $v_{\rm r}/v_{\rm K} \approx \alpha (a_{\rm disk}/v_{\rm K})^2$. Using $\dot{M}\approx 10^{-7}-10^{-6}~M_{\odot}~{\rm yr^{-1}}$ \citep{white2001,white2004} and $\alpha =0.01$ \citep{guilloteau2011}, we get $\rho_0\approx 10^{-9}-10^{-8}$~g~cm$^{-3}$. Here we adopt $\rho_0 = 10^{-8}$~g~cm$^{-3}$. 

From Equation~\ref{eq:disk_temp} and \ref{eq:disk_rho}, we obtain the magnetic field strength ($\approx B_{\rm \phi,disk}$) in the disk:
\begin{align}
B_{\rm \phi, disk} \approx 15~{\rm G}\left( \frac{\beta_{\rm disk}}{100}\right)^{-1/2}\left( \frac{r}{0.1~{\rm au}} \right)^{-21/16},\label{eq:disk_mag}
\end{align}
where $\beta_{\rm disk}$ is the plasma beta in the disk. We take $\beta_{\rm disk}=100$ as a fiducial value for MRI-active disks, following recent global MHD simulations \citep[e.g.][]{suzuki2014}.

We consider that the reconnection heating is balanced by the heat conduction cooling. The balance equation $E_{\rm H}=\partial F_{\rm hc}/\partial s $ gives
\begin{align}
f\frac{B_{\rm \phi,disk}^2}{8\pi}\omega_{\rm P} =  \frac{\kappa_0 T_{\rm c}^{7/2}}{\lambda_{\rm P}^2},\label{eq:balance}
\end{align}
where $T_{\rm c}$ is the disk coronal temperature. Using the expression of $\lambda_{\rm P} \approx 10 H_{\rm disk} = 10\sqrt{2} a_{\rm disk}/\Omega$, we obtain the average temperature from this relation:
\begin{align}
T_{\rm c} & \approx \left( \frac{40f}{\kappa_0 \gamma}\right)^{2/7} \beta_{\rm disk}^{-3/7}\Omega^{-2/7}a_{\rm disk}^{8/7}\rho_{\rm disk}^{2/7} \\
& \approx 3.4\times 10^6 ~{\rm K} \left( \frac{f}{0.1}\right)^{2/7} \left( \frac{M_*}{M_\odot}\right)^{-1/7}\left( \frac{\beta_{\rm disk}}{100}\right)^{-3/7} \left( \frac{r}{0.1~{\rm au}}\right)^{-15/28},\label{eq:corona_temp}
\end{align}
where $M_*$ is the stellar mass.

We estimate the density of the hot corona $\rho_{\rm c}$ by using the so-called RTV scaling law \citep{rosner1978}, which is derived from the energy balance between the heating and the radiative and the conductive cooling: $T_{\rm c}\approx c_{\rm RTV} (p_{\rm c} l)^{1/3}$ or equivalently
\begin{align}
\rho_{\rm c} \approx \mu c_{\rm RTV}^{-3} R_{\rm g}^{-1} T_{\rm c}^2 l^{-1},\label{eq:rtv}
\end{align}
where $c_{\rm RTV}$ is a constant determined by the atomic physics ($c_{\rm RTV}\approx 1.4\times 10^3$), $p_c=\rho_c R_{\rm g} T_{\rm c}/\mu$ is the coronal gas pressure, and $l$ is the loop length. Note that the RTV scaling is based on the atomic physics and the physics of heat conduction. Because of this universality, this relation has been applied to different systems such as the hot gas of the central regions of clusters of galaxies \citep{makishima2001}. Taking $l=\lambda_{\rm P}$ and using Equation~\ref{eq:corona_temp}, we obtain
\begin{align}
\rho_{\rm c} &\approx \frac{\mu}{10\sqrt{2}c_{\rm RTV}^3 R_g}\left(  \frac{40f}{\kappa_0 \gamma} \right)^{4/7}\beta_{\rm disk}^{-6/7}\Omega^{3/7}a_{\rm disk}^{9/7}\rho_{\rm disk}^{4/7} \\
& \approx 3.1 \times 10^{-17}~{\rm g~cm^{-3}}\left( \frac{f}{0.1}\right)^{4/7} \left( \frac{M_*}{M_\odot}\right)^{3/14}\left( \frac{\beta_{\rm disk}}{100}\right)^{-6/7} \left( \frac{r}{0.1~{\rm au}}\right)^{-123/56} \label{eq:corona_rho}
\end{align}
or
\begin{align}
n_{\rm c}& \approx 3.7 \times 10^{7}~{\rm cm^{-3}}\left( \frac{f}{0.1}\right)^{4/7} \left( \frac{M_*}{M_\odot}\right)^{3/14}\left( \frac{\beta_{\rm disk}}{100}\right)^{-6/7} \left( \frac{r}{0.1~{\rm au}}\right)^{-123/56} \label{eq:corona_den}
\end{align}
where the mean molecular weight $\mu$ is set to 0.5.

As we will see later, the plasma beta at the slow point $\beta_{\rm s}$ is a key quantity to determine the mass loss rate and the terminal velocity of the jet. We estimate the plasma beta as follows. Global MHD simulations by \citet{suzuki2014} suggest that the disk plasma beta and the ratio of the poloidal field strength $B_{\rm p}$ to the toroidal field strength $B_{\rm \phi}$ are almost constant in the disk with respect to the radius:
\begin{align}
\beta_{\rm disk} & ={\rm const.}\\
B_{\rm p}/B_{\phi} &\approx 0.1 {\it -} 0.3 \equiv f_{\rm B}
\end{align}
To obtain the coronal poloidal field strength, we assume the relation $B_{\rm p,s} \approx f_{\rm B} B_{\rm \phi,d}$. Using this relation and Equation~\ref{eq:disk_mag}, we get $\beta_{\rm s}$:
\begin{align}
\beta_{\rm s} \approx \frac{a_{\rm c}^2}{v_{\rm Ap,s}^2}
  \approx 3.8\times 10^{-3} \left( \frac{f_{\rm B}}{0.3}\right)^{-2} \left( \frac{M_*}{M_\odot}\right)^{-1/14} \left( \frac{\beta_{\rm disk}}{100}\right)^{-2/7}\left( \frac{r}{0.1~{\rm au}}\right)^{-3/28},\label{eq:beta_slow}
\end{align}
where $a_{\rm c}=\sqrt{\gamma R_{\rm g} T_{\rm c}}$ is the sound speed in the corona, and the specific heat ratio $\gamma$ is set to a lower value than the adiabatic one, 1.1, considering the heat conduction effect (although the result is not sensitive to the value). $v_{\rm Ap,s} = B_{\rm p}/\sqrt{4\pi \rho_{\rm s}}$ is the poloidal component of the Alfv\'en speed at the slow point.

\begin{figure}
\epsscale{.70}
\plotone{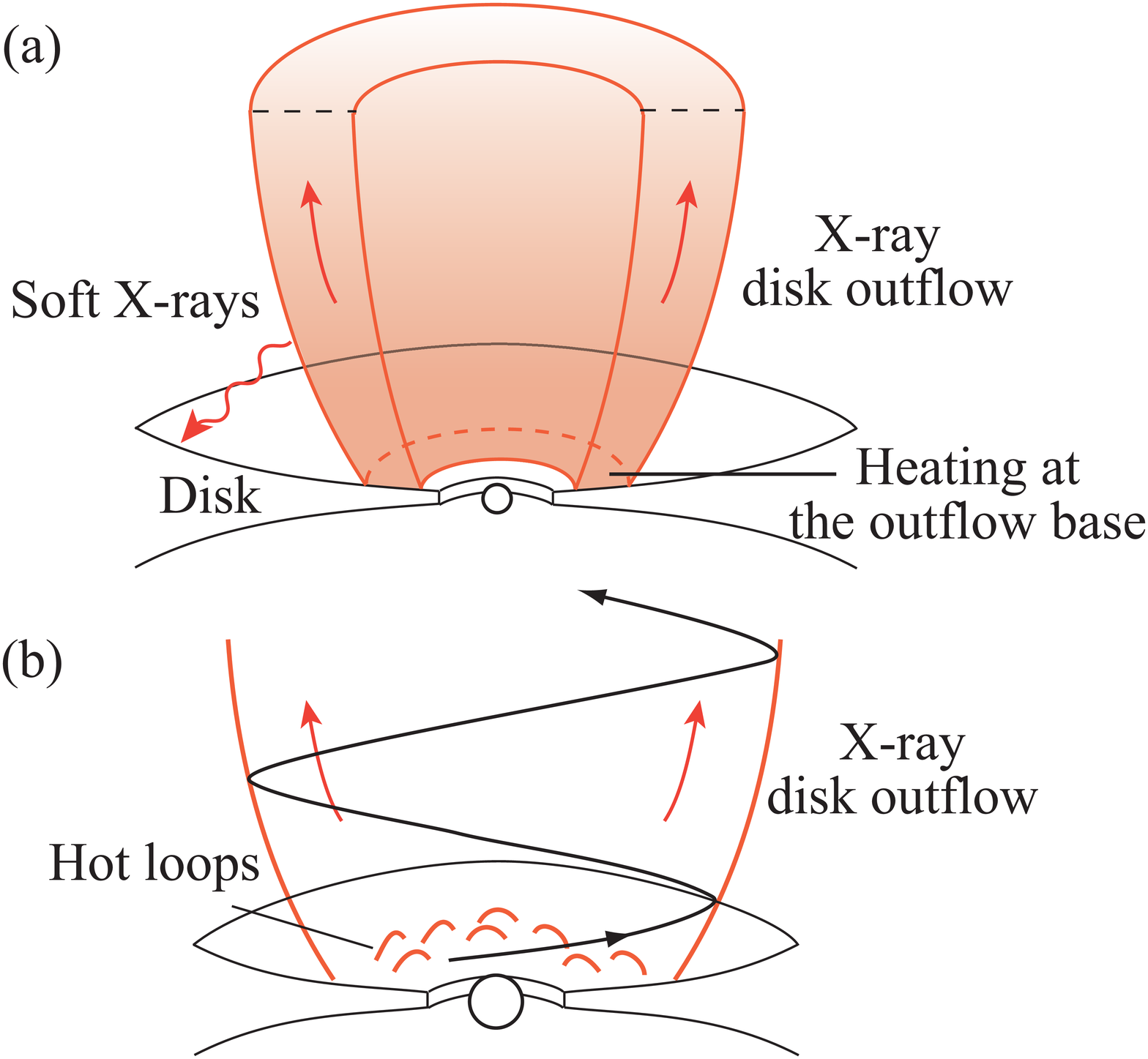}
\caption{A schematic cartoon of the X-ray disk jet. (a) Global picture of the X-ray disk jet model. (b) Enlarged picture of the central region. Many loops at the jet base denote magnetic loops formed as a result of the Parker instability in the disk. They are heated up by magnetic reconnection heating to form a hot disk corona. The X-ray jet emanates from the hot corona. \label{fig:jet}}
\end{figure}

A schematic cartoon of our model is presented in Figure~\ref{fig:jet}. Magnetic reconnection between emerging magnetic loops and ambient fields heats the disk atmosphere, forming a hot disk corona. The hot plasma flows out as X-ray jets. The driving mechanism of X-ray jets will be discussed later.

\section{Application to X-ray jets of DG Tau}

The temperature and mass loss rate of the X-ray jets from DG Tau were estimated from observations \citep{gudel2008, schneider2008}. We will compare these results with the quantities predicted by our disk corona model. The disk magnetic field strength which is required to sustain the hot corona will be investigated. 

\subsection{Temperature}\label{sec:temperature}
\citet{gudel2008} found that the X-ray jets and their base have the temperature of 3.4~MK. The disk coronal temperature predicted by our model is similar to the observed value (Equation~\ref{eq:corona_temp}). The temperature of the jet may decrease due to the expansion and radiative coolings, but these effects can be negligible under some reasonable conditions. 

Both the expansion and radiative coolings depend on the density and the opening angle of jets $\theta_{\rm op}$. Spatial resolution of observations is insufficient to determine the $\theta_{\rm op}$, and it is also difficult to theoretically predict the angle because the opening angle depends on many factors such as the disk magnetic field distribution. Here we examine whether there is a reasonable condition in which both effects can be negligible. Let us adopt $\theta_{\rm op}\approx 0.1$ and the jet speed $v\approx 3a_c\approx 10^8$~cm~s$^{-1}$ in the expanding jet region (see also the estimation of the jet speed in Section~\ref{sec:massloss}). Then, the density drops by $10^{-2}$ ($n\approx 4\times 10^5$~cm$^{-3}$ from Equation~\ref{eq:corona_den}) within $10^{6}$~s$\sim 0.03$~yr, which is much smaller than the radiative cooling time at the disk corona ($\sim 0.3$~yr). The radiative cooling with the density of $10^5$~cm$^{-3}$ becomes important on the timescale of a few years \citep{gudel2008}, which is comparable to the propagation timescale of the jet over the jet length (700~au). So if the density drops in the jet as fast as we assume here, we can neglect the radiative cooling effects.

The cooling by expansion can be estimated by $T^\prime/T_c = (r^\prime/r)^{2(\gamma_{\rm eff}-1)}$, where $T^\prime$ is the temperature in the expanded jet with the radius of $r^\prime$ \citep[see also][]{gudel2008}. $\gamma_{\rm eff}$ is the effective specific heat ratio, and can be close to unity due to the heat conduction. As a result of the expansion considered above, the temperature will be $T^\prime \simeq 0.8 T_c$ when $\gamma_{\rm eff}=1.05$. Therefore, if the opening angle is at such a value, then we may be able to neglect the expansion cooling.

\subsection{Mass Loss Rate}\label{sec:massloss}
The mass loss rate by the jet $\dot{M}$ can be expressed as $\dot{M} = (\rho A v)_{\rm s}$, where $\rho$ is the density, $A$ is the cross sectional area of the jet, and $v$ is the jet speed. The subscript $\rm s$ denotes the values at the slow point. The jet speed at the slow point $v_{\rm s}$ depends on the driving mechanisms: a general classification will be magnetically-driven or thermally-driven. The jet speed $v_{\rm s}$ is expressed as
\begin{align}
v_{\rm s}= 
\begin{cases}
v_{\rm slow,s} \hspace{2mm}(\rm magnetically \mbox{-}driven)\approx
      \left\{ 
        \begin{array}{lcl}
a_{\rm c}  & (B_{\rm \phi,s} \ll B_{\rm p,s}) \\
a_{\rm c} |B_{\rm p,s}/B_{\phi,s}| & (B_{\rm \phi,s} \gg B_{\rm p,s}) \\
        \end{array}
      \right. \\
      a_c \hspace{2mm}(\rm thermally \mbox{-}driven)
\end{cases}
\end{align}
where $v_{\rm slow,s}$ is the MHD slow-mode speed at the slow point, and $B_{\rm p,s}$ and $B_{\rm \phi,s}$ are the poloidal and azimuthal components of the magnetic fields at the slow point, respectively \citep[the expression for the magnetically-driven jets is from][]{kudoh1995,kudoh1997a}. Note that the velocity at the slow point is the projected slow-mode speed along a magnetic field to the poloidal plane, and the slow-mode speed in the low-beta plasma is similar to the sound speed. In the upper atmosphere, the magnetic pressure dominates the gas pressure because the gas pressure decreases more rapidly than the magnetic pressure with height (see also Equation~\ref{eq:beta_slow}). Therefore, we consider that the magnetic field at the slow point represents a coherent, nearly force-free magnetic field above the disk, rather than the disordered disk magnetic field in the MRI-turbulence.

We need to identify the main driving force of the X-ray jets in order to determine the scaling of the mass loss rate. While in realistic situations both magnetic and thermal mechanisms play a role in driving the jets, the terminal velocity is mainly determined by the dominant force(s). We aim to determine the dominant launching force by comparing the terminal velocities calculated from different forces. Following this strategy, we will investigate the mass loss rate later.

We can estimate the terminal velocity $v_\infty$ of magnetically-driven jets using the Michel's minimum energy solution (confirmed by \citet{kudoh1997a}):
\begin{align}
v_\infty = \left( \frac{\Omega^2 \Phi^2}{4\pi \dot{M}}\right)^{1/3},
\end{align}
where $\Phi = \pi B_{\rm p,s}r^2$ is the total magnetic flux and $\dot{M}$ is the mass loss rate of the jet \citep{michel1969}. $\dot{M}\approx (\rho Av)_{\rm s}$, where $\rho$ is the density, $A$ is the cross-sectional area of the jet, and $v$ is the jet speed.

Using the approximate relation $B_{\rm p,s}/B_{\rm \phi,s} \approx v_{\rm Ap,s}/v_{\rm K}$ (indicating a steady configuration of a magnetic field threading a Keplerian disk) and the plasma beta at the slow point $\beta_{\rm s}\approx (a_{\rm c}/v_{\rm Ap,s})^2$, we obtain
\begin{align}
v_{\infty} \approx
\begin{cases}
        \begin{array}{lcl}
\beta_s^{-1/3}(a_{\rm c}/v_{\rm K})^{1/3} v_{\rm K}&   (B_{\rm \phi,s} \ll B_{\rm p,s}) \\
\beta_s^{-1/6}v_{\rm K}  & (B_{\rm \phi,s} \gg B_{\rm p,s}) \\
        \end{array}\label{eq:v_inf}
\end{cases}
\end{align}
\citep[confirmed by 3D MHD simulations;][]{kigure2005}. The former is the terminal velocity of the magneto-centrifugal jets, and the latter is of the magnetic pressure gradient force driven jets. The terminal velocity in the case of the magneto-centrifugal acceleration will be
\begin{align}
v_\infty \approx 940~{\rm km~s}^{-1} \left( \frac{\beta_{\rm s}}{3\times 10^{-3}}\right)^{-1/3} \left( \frac{a_{\rm c}}{250~{\rm km~s}^{-1}}\right)^{1/3}  \left( \frac{v_{\rm K}}{100~{\rm km~s}^{-1}}\right)^{2/3},
\end{align}
which greatly exceeds (almost four times larger than) the coronal sound speed $a_{\rm c}\approx 250~{\rm km~s}^{-1}(T_{\rm c}/3.4\times 10^6~\rm K)^{0.5}$. On the other hand, the terminal velocity in the case of the magnetic pressure gradient force acceleration is comparable to $a_{\rm c}$:
\begin{align}
v_\infty \approx 260~{\rm km~s}^{-1} \left( \frac{\beta_{\rm s}}{3\times 10^{-3}}\right)^{-1/6}  \left( \frac{v_{\rm K}}{100~{\rm km~s}^{-1}}\right).
\end{align}
The above estimation indicates that the most important driving force will be the magneto-centrifugal force. However, the gas pressure may not be negligible because of its high temperature (the speed of the thermally driven jets can be larger than the coronal sound speed but should be of the order of it). For this reason, the X-ray jets is considered as a warm magneto-centrifugal jet.


We are ready to estimate the mass loss rate $\dot{M}$. The density structure in the sub slow-mode speed region (the region below the slow point) will be well approximated by the hydrostatic density structure as in the case of stellar winds \citep{lamers1999}. 
The lack of the information about detailed magnetic field structure prevents us from determining the slow point location. So for simplicity we take one pressure scale height from the disk as a typical slow point location. Then, we get $\rho_{\rm s}\approx \rho_{\rm c} \exp{(-1)}\approx 0.3\rho_{\rm c}$. Using $A\approx \pi r^2$ and $v_{\rm s}\approx a_{\rm c}$ (magneto-centrifugal jet), we obtain
\begin{align}
\dot{M} & = (\rho A v)_{\rm s}  \approx 0.3 \rho_c \cdot \pi r^2 \cdot a_{\rm c}\\
&\approx 2.5 \times 10^{-11} {\rm M_{\odot}~yr^{-1}} \left( \frac{f}{0.1}\right)^{6/7} \left( \frac{M_*}{M_\odot}\right)^{1/14} \left( \frac{\beta_{\rm disk}}{100}\right)^{-9/7} \left( \frac{r}{0.1~\rm au}\right)^{43/28},
\end{align}
which is similar to the observationally estimated mass loss rate \citep[$1.3\times 10^{-11}~\rm M_\odot~yr^{-1}$;][]{schneider2008}.

\subsection{Magnetic Field Strength in the Disk}
\begin{figure}
\epsscale{.70}
\plotone{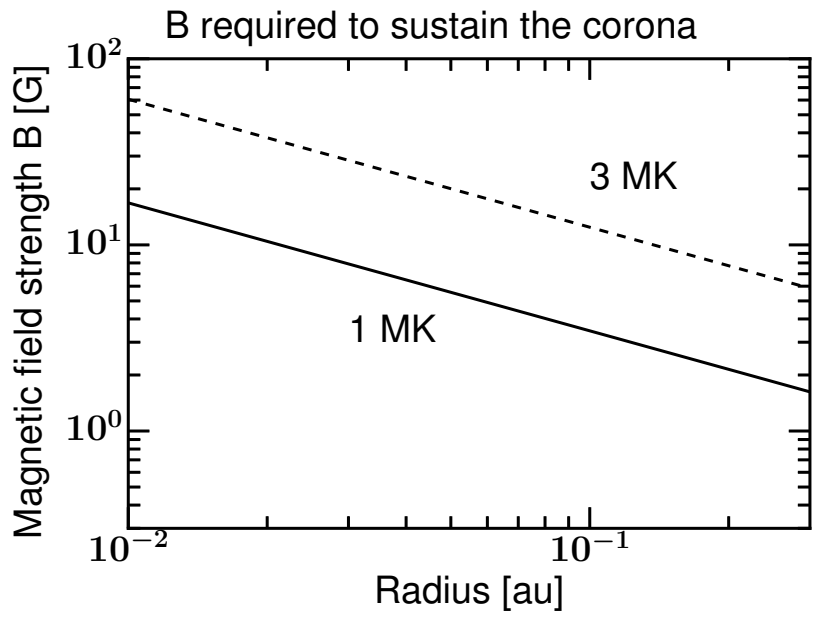}
\caption{Magnetic field strength required to sustain the 1~MK corona (solid) and 3~MK corona (dashed) as a function of radius. \label{fig:required_mag}}
\end{figure}

There is the threshold of the disk magnetic field strength required to sustain the MK-temperature corona. 
If we set $f=0.1$ and $M_*=M_\odot$, Equation~\ref{eq:corona_temp} gives
\begin{align}
\beta_{\rm disk} \approx 100 \left( \frac{T_{\rm c}}{3.4\times 10^6~\rm K}\right)^{-7/3} \left( \frac{r}{0.1~\rm au}\right)^{-5/4}\label{eq:disk_beta}
\end{align}
Using Equations~\ref{eq:disk_mag} and \ref{eq:disk_beta}, we calculate the threshold of the field strength. Figure~\ref{fig:required_mag} shows the thresholds of the field strength required to sustain the 1~MK (solid) and 3~MK (dashed) coronae. We found that only $3.5$~G is required at $r=0.1$~au to sustain a 1~MK corona.

The spatial resolution of the current observational instruments is insufficient to measure the field strength in the inner disk region. Our disk corona model could help indirectly estimating the field strength from the jet temperature $T_{\rm c}$ and the jet base radius. Using Equations~\ref{eq:disk_mag} and \ref{eq:corona_temp}, we can relate the field strength in the disk to the jet temperature:
\begin{align}
B_{\rm phi,disk} \approx 15~{\rm G}~\left( \frac{f}{0.1}\right)^{1/3}\left( \frac{T_{\rm c}}{3.4\times 10^6~\rm K}\right)^{7/6}\left( \frac{r}{0.1~\rm au}\right)^{-11/16}
\end{align}
where we set $M_*=M_\odot$. From this, the field strength in the disk is expected to be $\sim 15$~G.

\section{Effect of the X-ray from Jet on the Disk Dead-zone Size}
The X-ray photons from the jets will vertically enter the disk to increase the ionization degree, which can change the size of the dead-zone where MRI no longer operates. We investigate the effect of X-rays from the jets. 
For this aim, we adopt the Minimum Mass Solar Nebula (MMSN) model \citep{hayashi1981} as a typical outer disk model (observations indicate that the DG Tau disk is similar to the MMSN model; \citet{guilloteau2011}).
\begin{align}
T(r) & = 280~\left( \frac{r}{1~\rm au}\right)^{-1/2}~{\rm K}\\
\Sigma(r) & = 1.7\times 10^3 f_{\rm \Sigma}~\left( \frac{r}{1~\rm au}\right)^{-3/2}~{\rm g~cm^{-2}},
\end{align}
We set $f_{\rm \Sigma}$ to 1. Following \citet{guilloteau2011}, $T(1~{\rm au})\approx 300$~K and $f_{\Sigma}\approx 2$. However, for better comparison with other studies on the disk ionization degree, we take the above commonly-used parameter set \citep[e.g.][]{sano2000,mori2016}.

The ionization sources considered are cosmic rays, radioactive elements, stellar X-rays, and jet X-rays. The ionization by cosmic rays and radioactive elements are formulated in the same way as in \citet{sano2000}. The ionization by X-rays is calculated using the method given in \citet{glassgold1997} and \citet{fromang2002}. The recombination processes considered are the recombination on dust grains and radiative and dissociative recombination in the gas phase \citep[e.g.][]{umebayashi1980}. The charge reaction on dust grains mainly governs the charge neutrality \citep{umebayashi1983}. Using the relations among the number densities of electrons, molecular ions, and metal ions calculated by \citet{sano2000} (see Figure~4 in their paper), we simplify the charge reaction scheme \citep[see also][]{suzuki2010}. The dust grain size is assumed to be 0.1~$\rm \mu m$. The ionization degree is given by the steady solution of the rate equation for electrons.

The unresolved central point-spread function of DG Tau obtained by {\it Chandra} shows two separate thermal components: the hard and soft components. \citet{gudel2011} indicates that the soft component has almost the same absorption column density and the temperature as the those of a resolved X-ray jets, which suggests that the soft component originates in the unresolved inner jet. The luminosity of the soft-X-ray component is $L_{\rm X}\approx 10^{29}$~erg~s$^{-1}$, and this luminosity originates within the PSF with the spatial resolution of $\approx 0.2^{\prime\prime}\approx 30$~au.


There is uncertainty regarding the distance of the most X-ray luminous region in the inner jet from the central star. Therefore, approximating that the most luminous region is the point source, we change the distance $d_{\rm jet}$ of the source from the central star in the range of 0 to 30~au as a free parameter without changing the X-ray luminosity of $10^{29}$~erg~s$^{-1}$ and the energy of 0.4~keV \citep{gudel2008,gudel2011}. In this paper, we will show the three cases of $d_{\rm jet}=$3 (case~1), 10 (case~2), and 30~au (case~3). We take the case without the jet X-ray as the reference case.

Following \citet{gudel2008}, the stellar X-ray emission has the characteristic energy of 2~keV, and has the luminosity of $10^{30}$~erg~s$^{-1}$. The stellar X-ray source is located near the central star $(r,z)=(0, 0.05~\rm au)$ (i.e., 10~$R_\odot$ away from the star).

\begin{figure}
\epsscale{.70}
\plotone{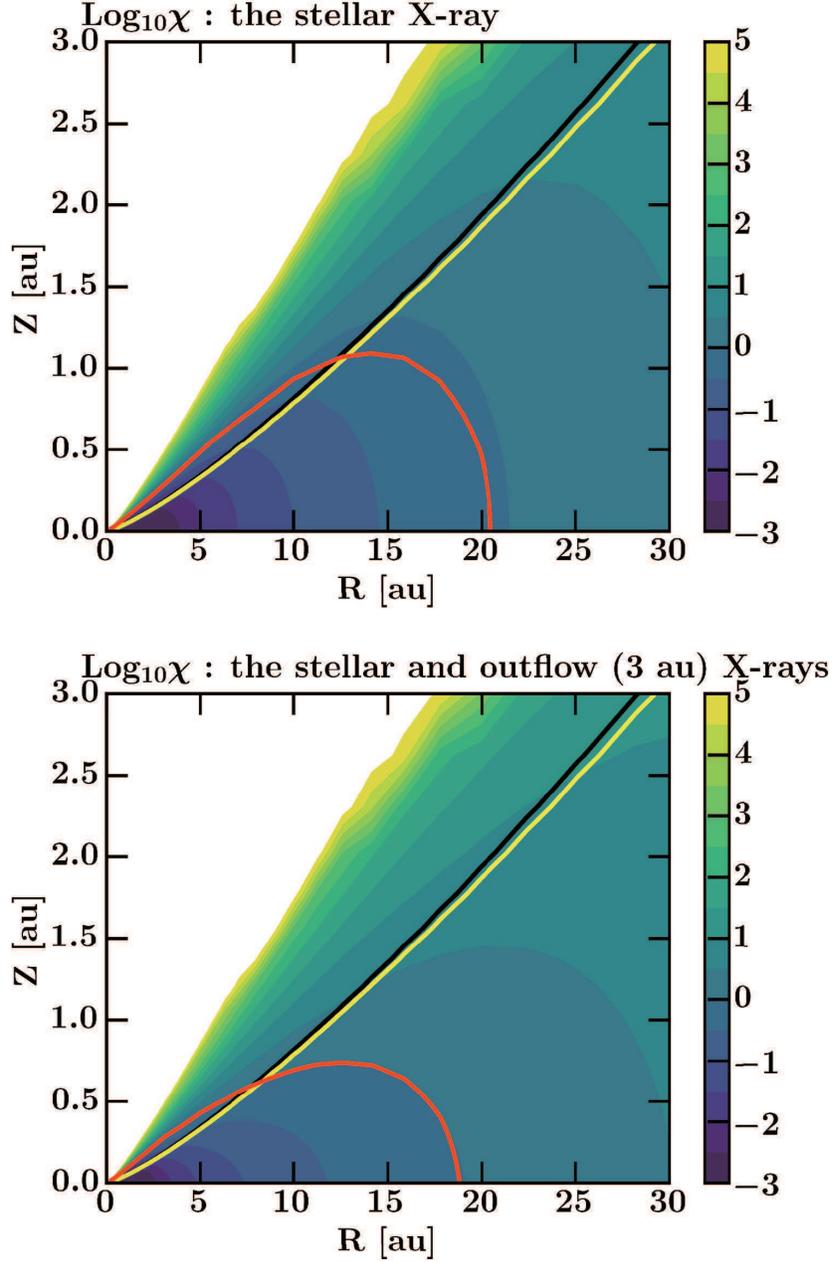}
\caption{Ionization degree distribution of the model disk. Top: The X-ray source is the stellar X-ray only. Bottom: The X-ray source is the combination of the stellar X-ray source and the jet X-ray source which is assumed to be located at $(r,z)=(0,3~\rm au)$. The ionization degree is normalized by $10^{-13}$. The black lines indicate the one pressure scale height. The yellow lines show the critical height of MRI for ideal MHD. The regions surrounded by the red line are the dead-zone. \label{fig:chi1}}
\end{figure}

Figure~\ref{fig:chi1} describes the dead-zone structure in the two different situations. The top panel shows the ionization degree (normalized by $10^{-13}$) of the reference case where the jet X-ray is not considered. The bottom panel exhibits the case with the jet X-ray from the distance of 3~au ($(r,z)=$(0, 3~au), case~1). The black lines denote the one pressure scale height. We introduce two parameters to indicate the MRI-active/inactive regions. One is the critical height $z_{\rm c}=\sqrt{\ln{\beta_{\rm disk}/4\pi^2}}H_{\rm disk}$ (yellow lines) which arises from the requirement that the characteristic MRI unstable wavelength in the ideal MHD limit should be smaller than the pressure scale height. MRI can only occur below this height. The other is the Els{\"a}sser number defined as $\Lambda_{\eta} = v_{{\rm A},z}^2/\eta\Omega $ (red lines), where $v_{{\rm A},z}$ is the Alfv\'en speed in the z-direction and $\eta$ is the resistivity. Here the resistivity is assumed to be the Ohmic resistivity only. The region below the line of $\Lambda_\eta=1$ is MRI-inactive and therefore the dead-zone. It is clearly shown that the dead-zone shrinks vertically in the case with the jet X-ray. The stellar X-ray cannot penetrate deeply in the radial direction because of the strong absorption by the dense gas near the central star, while the vertically injected X-ray from the jet can deeply penetrate in the disk in spite of the weaker and softer emission. This results in the shrinkage of the dead-zone.

\begin{figure}
\epsscale{.45}
\plotone{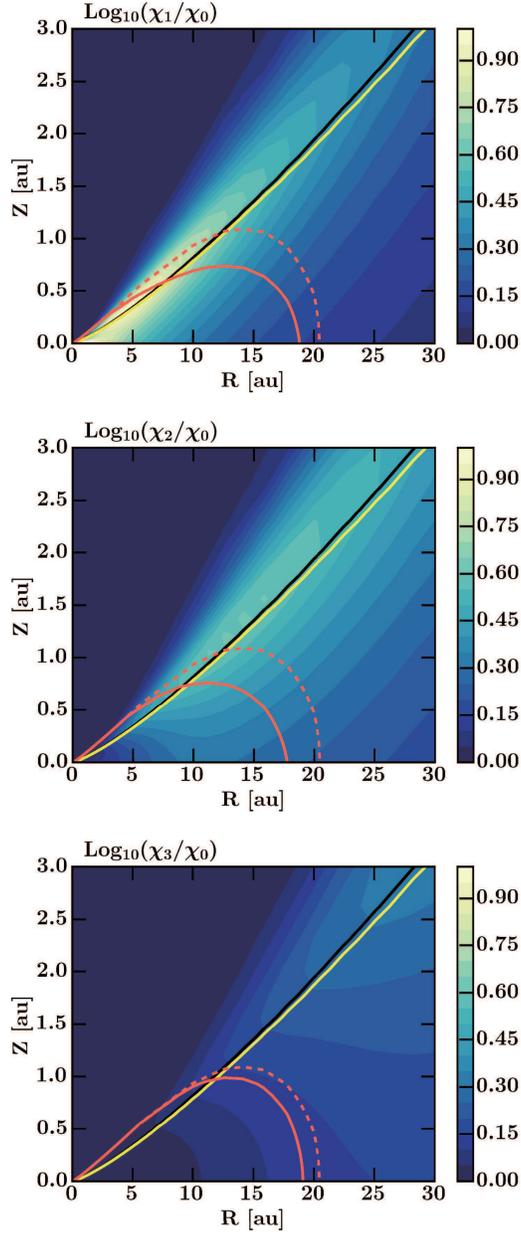}
\caption{Comparison of the ionization degree and the dead-zone size. Top: $\log_{10}{\chi_1/\chi_0}$, Middle: $\log_{10}{\chi_2/\chi_0}$, and Bottom: $\log_{10}{\chi_3/\chi_0}$, where $\chi_0$, $\chi_1$, $\chi_2$, and $\chi_3$ are the ionization degree of the reference case (no jet X-ray), case~1 (jet X-ray source at 3~au), case~2 (at 10~au), and case~3 (at 30~au), respectively. The black lines indicate the one pressure scale height. The yellow lines show the critical height of MRI for ideal MHD. The regions surrounded by the red line are the dead-zone. The dashed red lines are for the reference case. \label{fig:chi2}}
\end{figure}

Figure~\ref{fig:chi2} compares the three cases with different $d_{\rm jet}$. The color contour shows the ratio of the ionization degree of case~1 (Top), case~2 (Middle), and case~3 (Bottom) to that of the reference case. One will find a large increase of the ionization degree around the one pressure scale height. A clear enhancement near the mid-plane is also seen in the cases~1 and 2. In the case~2 the dead-zone shrinks in both the radial and vertical directions. In this case, the X-ray is injected more vertically than in the case~1, leading to the deeper X-ray penetration in the disk at a large ($>\sim 10$~au) radius. The dead-zone size does not change significantly in the case~3, because the larger distance makes the X-ray flux smaller than in the other cases. We note that the dead-zone of less than a few au does not change much in all the three cases, and even with larger plasma beta (we checked up to $10^4$).

\section{Summary and Discussion}

We presented a model of X-ray jets from young stellar objects, in which the disk atmosphere is heated by magnetic reconnection and the jet is launched mainly by the magneto-centrifugal force. Considering the energy balance between the heating and cooling at the base of the jets, we obtained the scaling relations for the temperature and density at the base, the mass loss rate, and the terminal velocity. We applied our ``disk corona model" to the X-ray jets of DG Tau, and found that this model can account for the observed temperature and estimated mass loss rate.

Our model can explain the presence of the puzzling stationary, steady, soft X-ray source at the jet base seen in the DG Tau jets \citep{gudel2008,gudel2011}. In the disk corona model, many small energy release events by magnetic reconnection heat the disk atmosphere repeatedly, which can be interpreted as a stationary and quasi-steady heating at the current spatial and temporal resolutions of observations. In addition, because the jet in our model is originated not from the stellar magnetosphere but from the disk atmosphere, the temporal behavior of the jet will be independent of the stellar flaring activity. This prediction is consistent with the observations. The X-ray emitting component which could show a proper motion may require different heating mechanisms such as shock heating. Our model may account for the stationary X-ray source of L1551 IRS5 jet (HH154), although we should note that the source we can observe is located at $\sim$100~au and the very footpoint of the jet is not seen due to a strong absorption. 
As discussed in Section~\ref{sec:temperature}, magnetically heated jets retain its high temperature even at more than 100~au if the jets have a sufficiently high speed and do not experience a severe cooling. Considering this, we infer that L1551 IRS5 jet may be an example of the magnetically heated jets.

We investigated the influence of the soft X-ray from jets on the dead-zone size in the disk on the basis of the observational constraints. We found that the dead-zone size can be significantly smaller when we include the jet X-ray source within the distance of $\sim$10~au from the central star. In the disk corona model, the jet base will be the brightest region because of its high density. The pressure scale height, which will be a good measure of the scale of the jet base, is $\approx 1~{\rm au}~(T/3.4~{\rm MK})^{1/2}(r/0.1~{\rm au})^{3/2}$. Therefore, the jet base will be the most X-ray bright region, and can significantly affect the disk ionization degree because it is close to the disk. For this reason, we consider that it will be important to include the X-ray emission from jets as an additional ionization source in the disk models. The jet X-ray may affect the influences of non-ideal MHD effects near the disk surface through the ionization degree \citep[e.g.][]{bai2011}, although the effect seems small within a few au.

Our model suggests that we can estimate the disk magnetic field strength from the jet temperature. The magnetic field strength is an important quantity for determining the jet launching mechanism and accretion rate in the disk, but in general it is difficult to directly measure the field strength. Our model could provide a useful tool to estimate the field strength in the inner disk from the jet temperature and the jet base radius.

It is possible that the temperature of X-ray jets exceeds the virial temperature at the launching radius when the heating process is drastic and/or the hot plasma is confined in magnetic loops. This is supported from the fact that the active stars retain (magnetically heated) hot coronae with the temperature much larger than their virial temperature \citep[e.g.][]{gudel2004}. However, we should note that the heating process and magnetic confinement depends on the detailed magnetic field configuration. For instance, near the inner edge of the disk, the interaction between the stellar magnetosphere and the disk can lead to complex magnetic and flow structures \citep{shu1994,hayashi1996,romanova2009}, which makes it difficult to infer the jet structure in the proximity of the inner disk. A detailed modeling of the dynamic atmospheric structure is required for drawing a definitive conclusion.

\acknowledgments
We thank S. Okuzumi and M. Kunitomo for encouraging the investigation of the effect of the jet X-rays on the ionization degree of the disk. ST acknowledges support by the Research Fellowship of the Japan Society for the Promotion of Science (JSPS). This work is supported in part by Grants-in-Aid for Scientific Research from the MEXT of Japan, 17H01105 (TKS).

\begin{thebibliography}{}
\bibitem[Akeson et al.(2005)]{akeson2005} Akeson, R.~L., Boden, A.~F., Monnier, J.~D., et al.\ 2005, \apj, 635, 1173 
\bibitem[Bacciotti et al.(2000)]{bacciotti2000} Bacciotti, F., Mundt, R., Ray, T.~P., et al.\ 2000, \apjl, 537, L49
\bibitem[Bai(2011)]{bai2011} Bai, X.-N.\ 2011, \apj, 739, 50 
\bibitem[Bally et al.(2003)]{bally2003} Bally, J., Feigelson, E., \& Reipurth, B.\ 2003, \apj, 584, 843 
\bibitem[Blandford \& Payne(1982)]{blandford1982} Blandford, R.~D., \& Payne, D.~G.\ 1982, \mnras, 199, 883 
\bibitem[Bonito et al.(2010)]{bonito2010} Bonito, R., Orlando, S., Miceli, M., et al.\ 2010, \aap, 517, A68 
\bibitem[Bonito et al.(2011)]{bonito2011} Bonito, R., Orlando, S., Miceli, M., et al.\ 2011, \apj, 737, 54 
\bibitem[Bouvier et al.(1993)]{bouvier1993} Bouvier, J., Cabrit, S., Fernandez, M., Martin, E.~L., \& Matthews, J.~M.\ 1993, \aap, 272, 176 
\bibitem[Cirtain et al.(2013)]{cirtain2013} Cirtain, J.~W., Golub, L., Winebarger, A.~R., et al.\ 2013, \nat, 493, 501 
\bibitem[Dullemond et al.(2007)]{dullemond2007} Dullemond, C.~P., Hollenbach, D., Kamp, I., \& D'Alessio, P.\ 2007, Protostars and Planets V, 555 
\bibitem[Ercolano et al.(2009)]{ercolano2009} Ercolano, B., Clarke, C.~J., \& Drake, J.~J.\ 2009, \apj, 699, 1639 
\bibitem[Favata et al.(2002)]{favata2002} Favata, F., Fridlund, C.~V.~M., Micela, G., Sciortino, S., \& Kaas, A.~A.\ 2002, \aap, 386, 204 
\bibitem[Frank et al.(2014)]{frank2014} Frank, A., Ray, T.~P., Cabrit, S., et al.\ 2014, Protostars and Planets VI, 451 
\bibitem[Fridlund et al.(2005)]{fridlund2005} Fridlund, C.~V.~M., Liseau, R., Djupvik, A.~A., et al.\ 2005, \aap, 436, 983 
\bibitem[Fromang et al.(2002)]{fromang2002} Fromang, S., Terquem, C., \& Balbus, S.~A.\ 2002, \mnras, 329, 18 
\bibitem[Glassgold et al.(1997)]{glassgold1997} Glassgold, A.~E., Najita, J., \& Igea, J.\ 1997, \apj, 480, 344 
\bibitem[Gorti et al.(2009)]{gorti2009} Gorti, U., Dullemond, C.~P., \& Hollenbach, D.\ 2009, \apj, 705, 1237 
\bibitem[G{\"u}del(2004)]{gudel2004} G{\"u}del, M.\ 2004, \aapr, 12, 71 
\bibitem[G{\"u}del et al.(2005)]{gudel2005} G{\"u}del, M., Skinner, S.~L., Briggs, K.~R., et al.\ 2005, \apjl, 626, L53 
\bibitem[G{\"u}del et al.(2007)]{gudel2007} G{\"u}del, M., Telleschi, A., audard, M., et al.\ 2007, \aap, 468, 515 
\bibitem[G{\"u}del et al.(2008)]{gudel2008} G{\"u}del, M., Skinner, S.~L., audard, M., Briggs, K.~R., \& Cabrit, S.\ 2008, \aap, 478, 797 
\bibitem[G{\"u}del et al.(2011)]{gudel2011} G{\"u}del, M., audard, M., Bacciotti, F., et al.\ 2011, 16th Cambridge Workshop on Cool Stars, Stellar Systems, and the Sun, 448, 617 
\bibitem[Guilloteau et al.(2011)]{guilloteau2011} Guilloteau, S., Dutrey, A., Pi{\'e}tu, V., \& Boehler, Y.\ 2011, \aap, 529, A105 
\bibitem[G{\"u}nther et al.(2014)]{gunther2014} G{\"u}nther, H.~M., Li, Z.-Y., \& Schneider, P.~C.\ 2014, \apj, 795, 51 
\bibitem[Hawley et al.(1995)]{hawley1995} Hawley, J.~F., Gammie, C.~F., \& Balbus, S.~A.\ 1995, \apj, 440, 742 
\bibitem[Hayashi(1981)]{hayashi1981} Hayashi, C.\ 1981, Progress of Theoretical Physics Supplement, 70, 35 
\bibitem[Hayashi et al.(1996)]{hayashi1996} Hayashi, M.~R., Shibata, K., \& Matsumoto, R.\ 1996, \apjl, 468, L37
\bibitem[Hirose \& Turner(2011)]{hirose2011} Hirose, S., \& Turner, N.~J.\ 2011, \apjl, 732, L30 
\bibitem[Io \& Suzuki(2014)]{io2014} Io, Y., \& Suzuki, T.~K.\ 2014, \apj, 780, 46 
\bibitem[Kigure \& Shibata(2005)]{kigure2005} Kigure, H., \& Shibata, K.\ 2005, \apj, 634, 879 
\bibitem[Kudoh \& Shibata(1995)]{kudoh1995} Kudoh, T., \& Shibata, K.\ 1995, \apjl, 452, L41 
\bibitem[Kudoh \& Shibata(1997)]{kudoh1997a} Kudoh, T., \& Shibata, K.\ 1997, \apj, 474, 362 
\bibitem[Kusaka et al.(1970)]{kusaka1970} Kusaka, T., Nakano, T., \& Hayashi, C.\ 1970, Progress of Theoretical Physics, 44, 1580 
\bibitem[Lamers \& Cassinelli(1999)]{lamers1999} Lamers, H.~J.~G.~L.~M., \& Cassinelli, J.~P.\ 1999, Introduction to Stellar Winds, by Henny J.~G.~L.~M.~Lamers and Joseph P.~Cassinelli, pp.~452.~ISBN 0521593980.~Cambridge, UK: Cambridge University Press, June 1999., 452 
\bibitem[Lavalley-Fouquet et al.(2000)]{lavalley2000} Lavalley-Fouquet, C., Cabrit, S., \& Dougados, C.\ 2000, \aap, 356, L41 
\bibitem[Makishima et al.(2001)]{makishima2001} Makishima, K., Ezawa, H., Fukuzawa, Y., et al.\ 2001, \pasj, 53, 401 
\bibitem[Matsumoto \& Tajima(1995)]{matsumoto1995} Matsumoto, R., \& Tajima, T.\ 1995, \apj, 445, 767 
\bibitem[Michel(1969)]{michel1969} Michel, F.~C.\ 1969, \apj, 158, 727
\bibitem[Mori \& Okuzumi(2016)]{mori2016} Mori, S., \& Okuzumi, S.\ 2016, \apj, 817, 52 
\bibitem[Parker(1966)]{parker1966} Parker, E.~N.\ 1966, \apj, 145, 811 
\bibitem[Parker(1988)]{parker1988} Parker, E.~N.\ 1988, \apj, 330, 474 
\bibitem[Pyo et al.(2009)]{pyo2009} Pyo, T.-S., Hayashi, M., Kobayashi, N., Terada, H., \& Tokunaga, A.~T.\ 2009, \apj, 694, 654 
\bibitem[Pravdo et al.(2001)]{pravdo2001} Pravdo, S.~H., Feigelson, E.~D., Garmire, G., et al.\ 2001, \nat, 413, 708 
\bibitem[Pudritz \& Norman(1986)]{pudritz1986} Pudritz, R.~E., \& Norman, C.~A.\ 1986, \apj, 301, 571 
\bibitem[Raga et al.(2002)]{raga2002} Raga, A.~C., Noriega-Crespo, A., \& Vel{\'a}zquez, P.~F.\ 2002, \apjl, 576, L149 
\bibitem[Ray et al.(2007)]{ray2007} Ray, T., Dougados, C., Bacciotti, F., Eisl{\"o}ffel, J., \& Chrysostomou, A.\ 2007, Protostars and Planets V, 231 
\bibitem[Rodr{\'{\i}}guez et al.(2012)]{rodriguez2012} Rodr{\'{\i}}guez, L.~F., Gonz{\'a}lez, R.~F., Raga, A.~C., et al.\ 2012, \aap, 537, A123 
\bibitem[Romanova et al.(2009)]{romanova2009} Romanova, M.~M., Ustyugova, G.~V., Koldoba, A.~V., \& Lovelace, R.~V.~E.\ 2009, \mnras, 399, 1802 
\bibitem[Rosner et al.(1978)]{rosner1978} Rosner, R., Tucker, W.~H., \& Vaiana, G.~S.\ 1978, \apj, 220, 643 
\bibitem[Sakurai(1985)]{sakurai1985} Sakurai, T.\ 1985, \aap, 152, 121 
\bibitem[Sano et al.(2000)]{sano2000} Sano, T., Miyama, S.~M., Umebayashi, T., \& Nakano, T.\ 2000, \apj, 543, 486 
\bibitem[Sano \& Inutsuka(2001)]{sano2001} Sano, T., \& Inutsuka, S.-i.\ 2001, \apjl, 561, L179 
\bibitem[Schneider \& Schmitt(2008)]{schneider2008} Schneider, P.~C., \& Schmitt, J.~H.~M.~M.\ 2008, \aap, 488, L13 
\bibitem[Schneider et al.(2011)]{schneider2011} Schneider, P.~C., G{\"u}nther, H.~M., \& Schmitt, J.~H.~M.~M.\ 2011, \aap, 530, A123 
\bibitem[Schneider et al.(2013a)]{schneider2013} Schneider, P.~C., Eisl{\"o}ffel, J., G{\"u}del, M., et al.\ 2013, \aap, 550, L1 
\bibitem[Schneider et al.(2013b)]{schneider2013b} Schneider, P.~C., Eisl{\"o}ffel, J., G{\"u}del, M., et al.\ 2013, \aap, 557, A110 
\bibitem[Schwenn(2006)]{schwenn2006} Schwenn, R.\ 2006, Living Reviews in Solar Physics, 3,  
\bibitem[Shibata \& Uchida(1986)]{shibata1986} Shibata, K., \& Uchida, Y.\ 1986, \pasj, 38, 631 
\bibitem[Shu et al.(1994)]{shu1994} Shu, F., Najita, J., Ostriker, E., et al.\ 1994, \apj, 429, 781 
\bibitem[Skinner \& G{\"u}del(2014)]{skinner2014} Skinner, S.~L., \& G{\"u}del, M.\ 2014, \apj, 788, 101
\bibitem[Skinner et al.(2016)]{skinner2016} Skinner, S.~L., audard, M., \& G{\"u}del, M.\ 2016, \apj, 826, 84 
\bibitem[Spitzer(1962)]{spitzer1962} Spitzer, L.\ 1962, Physics of Fully Ionized Gases, New York: Interscience (2nd edition), 1962
\bibitem[Staff et al.(2010)]{staff2010} Staff, J.~E., Niebergal, B.~P., Ouyed, R., Pudritz, R.~E., \& Cai, K.\ 2010, \apj, 722, 1325 
\bibitem[Suzuki et al.(2010)]{suzuki2010} Suzuki, T.~K., Muto, T., \& Inutsuka, S.-i.\ 2010, \apj, 718, 1289 
\bibitem[Suzuki \& Inutsuka(2014)]{suzuki2014} Suzuki, T.~K., \& Inutsuka, S.-i.\ 2014, \apj, 784, 121 
\bibitem[Takahara et al.(1989)]{takahara1989} Takahara, F., Rosner, R., \& Kusunose, M.\ 1989, \apj, 346, 122 
\bibitem[Turner \& Sano(2008)]{turner2008} Turner, N.~J., \& Sano, T.\ 2008, \apjl, 679, L131 
\bibitem[Umebayashi \& Nakano(1980)]{umebayashi1980} Umebayashi, T., \& Nakano, T.\ 1980, \pasj, 32, 405 
\bibitem[Umebayashi(1983)]{umebayashi1983} Umebayashi, T.\ 1983, Progress of Theoretical Physics, 69, 480 
\bibitem[Ustyugova et al.(1999)]{ustyugova1999} Ustyugova, G.~V., Koldoba, A.~V., Romanova, M.~M., Chechetkin, V.~M., \& Lovelace, R.~V.~E.\ 1999, \apj, 516, 221 
\bibitem[White \& Ghez(2001)]{white2001} White, R.~J., \& Ghez, A.~M.\ 2001, \apj, 556, 265
\bibitem[White \& Hillenbrand(2004)]{white2004} White, R.~J., \& Hillenbrand, L.~A.\ 2004, \apj, 616, 998 
\end{thebibliography}
\end{document}